\documentclass[aps,pre,twocolumn]{revtex4-1}
\usepackage{graphics,graphicx,epsfig}
\usepackage{amssymb,color}
\usepackage{epsf,epstopdf,wrapfig}

\usepackage{graphics,graphicx,epsfig}
\usepackage{amssymb,color}
\usepackage{epsf,epstopdf,wrapfig}

\begin{document}

\title{Searching for simplicity: Approaches to the analysis of neurons and behavior}

\author{Greg J. Stephens,$^{a,b}$ Leslie C. Osborne$^d$ and William Bialek$^{a,b,c}$}

\affiliation{$^a$Joseph Henry Laboratories of Physics,
$^b$Lewis--Sigler Institute for Integrative Genomics,
and $^c$Princeton Center for Theoretical Science\\
Princeton University, Princeton, New Jersey 08544 USA\\
$^d$Department of Neurobiology, University of Chicago, Chicago, Illinois 60637}

\begin{abstract} 
What fascinates us about animal behavior is its richness and complexity, but {\em understanding} behavior and its neural basis requires a simpler description. Traditionally, simplification has been imposed by training animals to engage in a limited set of behaviors, by hand scoring behaviors into discrete classes, or by limiting the sensory experience of the organism.   An alternative is to ask whether we can search through the dynamics of natural behaviors to find explicit evidence that these behaviors are simpler than they might have been.  We review two mathematical approaches to simplification, dimensionality reduction and the maximum entropy method, and we draw on examples from different levels of biological organization, from the crawling behavior of {\em C. elegans} to the control of smooth pursuit eye movements in primates, and from the coding of natural scenes by networks of neurons in the retina to the rules of English spelling.  In each case, we argue that the explicit search for simplicity  uncovers new and unexpected features of the biological system, and that the evidence for simplification gives us a language with which to  phrase new questions for the next generation of experiments. The fact that similar mathematical structures succeed in taming the complexity of very different biological systems hints that there is something more general to be discovered.
 \end{abstract}
 
 
 \maketitle

\section{Introduction}

The last decades have seen an explosion in our ability to characterize the microscopic mechanisms---the molecules, cells, and circuits---that generate the behavior of biological systems.  In contrast, our characterization of behavior itself has advanced much more slowly.  Starting in the late nineteenth century, attempts to quantify behavior focused on experiments in which the behavior itself was restricted, for example by forcing an observer to choose among a limited set of alternatives.  In the mid--twentieth century, ethologists emphasized the importance of observing behavior in its natural context, but here, too, the analysis most often focused on the counting of discrete actions.  Parallel to these efforts, neurophysiologists were making progress on how the brain represents the sensory world by presenting simplified stimuli and labeling cells by preference for stimulus features.

Here we outline an approach in which living systems naturally explore a relatively unrestricted space of motor outputs or neural representations, and we search directly for simplification within the data.
While there is often suspicion of attempts to reduce the evident complexity of the brain, 
it is unlikely that understanding will be achieved without some sort of  compression.  Rather than restricting behavior (or our description of behavior) from the outset, we will let the system ``tell us'' whether our favorite simplifications are successful.  Furthermore, 
we start with high spatial and temporal resolution data since we do not know the simple representation ahead of time.  
This approach is made possible only by the combination of new experimental methods that generate  larger, higher quality data sets with  the application of mathematical ideas that have a chance of discovering unexpected simplicity in these complex systems. We present four very different examples where finding such simplicity informs our understanding of biological function.

\section{Dimensionality reduction}

In the human body there are approximately 100 joint angles and substantially more muscles.   Even if  each muscle has just two states (rest or tension),    the number of possible postures is enormous, $2^{N_{\rm muscles}} \sim 10^{30}$.  If our bodies moved aimlessly among these states, characterizing our motor behavior would be hopeless---no  experiment could sample even a tiny fraction of all the possible trajectories.  Moreover, wandering in a high dimensional space is unlikely to generate functional actions that make sense in a realistic context.  Indeed, it is doubtful that a plausible neural system would independently control all the muscles and joint angles without some coordinating patterns or ``movement primatives" from which to build a repertoire of actions.  There have been several motor systems in which just such a reduction in dimensionality has been found \cite{nelson_83,davella+bizzi_98,santello+al_98,sanger_00,ingram+al_08}.  Here we present two examples of behavioral dimensionality reduction which represent very different levels of system complexity: smooth pursuit eye movements in monkeys and the free wiggling of worm-like nematodes. These examples are especially compelling as so few dimensions are required for a complete description of natural behavior.
\subsection*{Smooth pursuit eye movements}

Movements are variable even if conditions are carefully repeated, but the origin of that variability is poorly understood.  Variation might arise from noise in sensory processing to identify goals for movement, in planning or generating movement commands, or in the mechanical response of the muscles.   The structure of behavioral variation can inform our understanding of the underlying system if we can connect the dimensions of variation to a particular stage of neural processing. 

 \begin{figure}[ht]
\includegraphics[width=0.85\columnwidth]{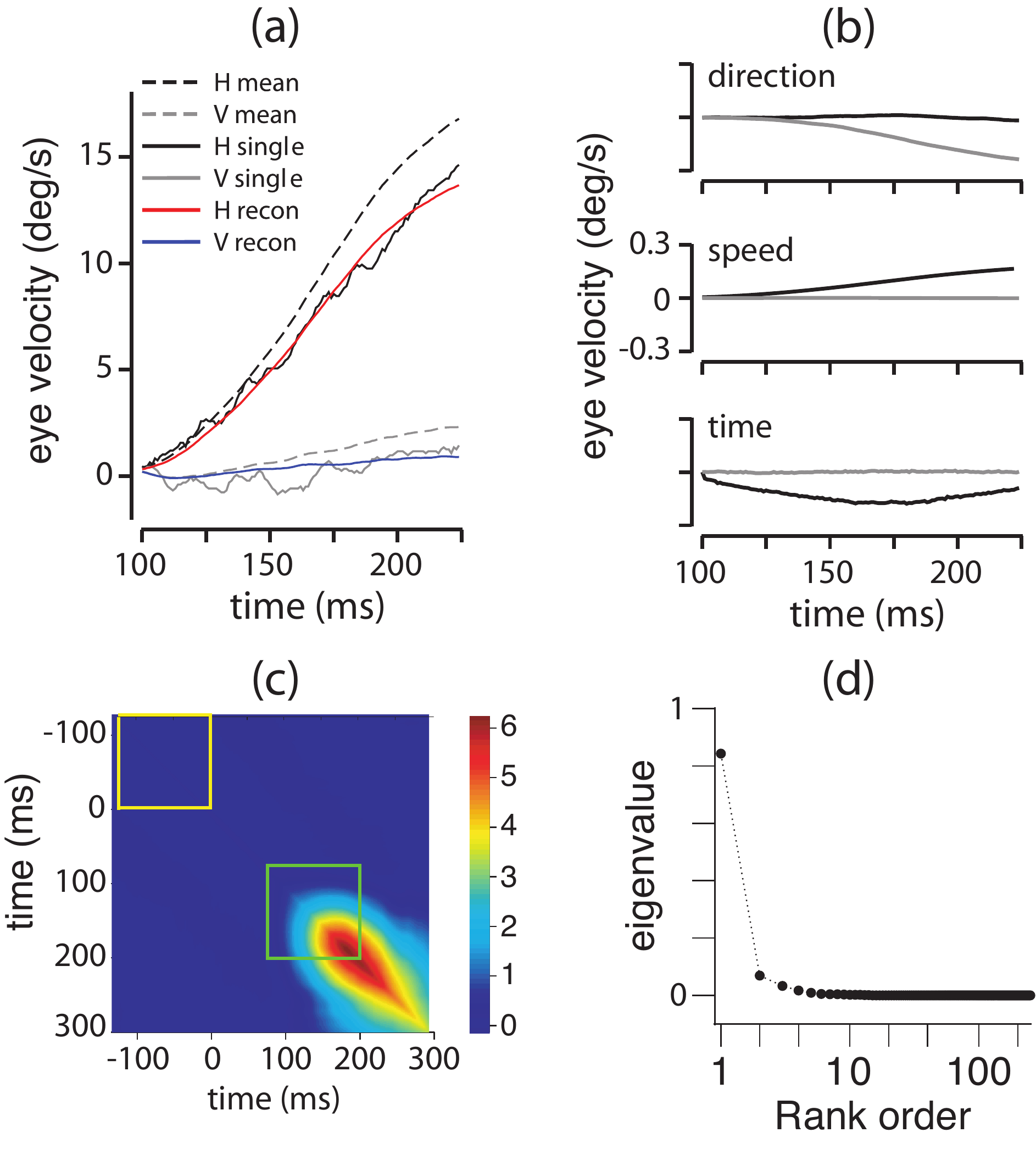}
\caption{The low-dimensional dynamics of  pursuit eye velocity trajectories \cite{osborne+al_05}. (a) Eye movements were recorded from male rhesus monkeys ({\it Macaca mulatta}) that had been trained to fixate and track visual targets.  Thin black and gray lines represent H and V eye velocity in response to a step in target motion on a single trial; dashed lines represent the corresponding trial-averaged means.  Red and blue lines represent the model prediction (b) Three natural modes of variation corresponding to direction, speed and time provide an essentially complete basis for eye trajectories. Black and gray curves correspond to H and V components (c) The covariance matrix of the horizontal eye velocity trajectories.  The yellow square marks 125ms during the fixation period prior to target motion onset, the green square the first 125ms of pursuit.  The color scale is in $\mathrm{deg/s}^2$ (d)  The eigenvalue spectrum of the difference matrix $\Delta C (t,t')=C_{\rm pursuit} (t,t') {\rm (green \,square)}-C_{\rm background} (t,t') {\rm (yellow \,square)}$. }
\label{fig:fig1}
\end{figure}

Like other types of movement, eye movements are potentially high dimensional if eye position and velocity vary independently from moment to moment.   But an analysis of the  natural variation in smooth pursuit eye movement behavior reveals a simple structure whose form suggests a neural origin for the noise that gives rise to behavioral variation.  Pursuit is a tracking eye movement, triggered by image motion on the retina, which serves to stabilize a target's retinal image and thus to prevent motion blur \cite{rashbass_61}.    When a target begins to move relative to the eye, the pursuit system interprets the resulting image motion on the retina to estimate the target's trajectory and then to  accelerate the eye to match the target's motion direction and speed.  While tracking on longer time scales is driven by both retinal inputs and by extra--retinal feedback signals, the initial $\sim 125\,{\rm ms}$ of the movement is generated purely from sensory estimates of the target's motion, using visual inputs present before the onset of the response.  Focusing on just this initial portion of the pursuit movement, we can express the eye velocity in response to steps in target motion as a vector, $\mathbf{v}(t)= v_H(t) \mathbf{\hat{i}} + v_V(t) \mathbf{\hat{j}} $, where  $v_H(t)$ and $v_V(t)$ are the horizontal and vertical components of the velocity, respectively (solid black and gray lines in Fig 1a).  If the initial $125\,{\rm ms}$ of eye movement is sampled every millisecond, the pursuit trajectories have 250 dimensions.

We compute the covariance of fluctuations about the mean trajectory, shown in Fig \ref{fig:fig1}c. Focusing on a window of  $125\,{\rm ms}$ at the start of the pursuit response (green box), we find that the first three eigenvalues of the covariance matrix are larger than the rest, which we confirmed by estimating the standard error of the values for each dataset \cite{osborne+al_05}.  This low dimensional structure is not a limitation of the motor system, since during fixation (yellow box) there are 80 significant eigenvalues.  Indeed, the small amplitude, high dimensional variation visible during  fixation appears to be an ever present background noise that is swamped by the larger fluctuations in movement specific to pursuit.  If the covariance of this background noise  is subtracted from the covariance during pursuit, the 3 dimensional structure becomes essentially exact,  accounting for $\sim 94\%$ of variations in eye velocity. 

How does low dimensionality in eye movement arise?  The goal of the movement is to match the eye to the target's velocity, which is constant in these experiments. The brain must therefore interpret the activity of sensory neurons that represent its visual inputs,  detecting that the target has begun to move (at time $t_0$) and estimating the direction $\theta$ and speed $v$ of motion.   At best, the brain estimates these quantities and transforms these estimates into some desired trajectory of eye movements, which we can write as $\mathbf{v}(t; \hat t_0 , \hat\theta , \hat v)$, where $\hat{\cdot}$ denotes an estimate of the quantity $\cdot$.  But estimates are never perfect, so we should imagine that $\hat t_0 = t_0 +\delta t_0$, and so on, where $\delta t_0$ is the small error in the sensory estimate of target motion onset on a single trial.  If these errors are small, we can write
\begin{eqnarray}
\mathbf{v}(t) &=& \mathbf{v}(t;t_0,v,\theta) +   \delta t_0  \frac{\partial \mathbf{v}(t;t_0,v,\theta)}{\partial t_0} + \delta \theta  \frac{\partial \mathbf{v}(t;t_0,v,\theta)}{\partial \theta} \nonumber \\
&&+ \delta v  \frac{\partial \mathbf{v}(t;t_0,v,\theta)}{\partial v} +\delta\mathbf{v}_{\mathrm{back}}(t),
\label{decomp}
\end{eqnarray}
where the first term is the average eye movement made in response many repetitions of the target motion, the next three terms describe the effects of the sensory errors, and the final term is the background noise.  Thus, if we can separate out the effects of the background noise, the fluctuations in $\bf v(t)$ from trial to trial should be described by just three random numbers, $\delta t_0$, $\delta \theta$, and $\delta v$:  the variations should be three dimensional, as observed.

The partial derivatives in Eq (\ref{decomp}) can be measured as the difference between the trial-averaged pursuit trajectories in response to slightly different target motions.  In fact the average trajectories vary in a simple way, shifting along the $t$ axis as we change $t_0$, rotating in space as we change $\theta$, and scaling uniformly faster or slower as we change $v$ \cite{osborne+al_05}, so that the relevant derivatives can be estimated just from one average trajectory.  When the dust settles, this means that we can write the covariance of fluctuations around the mean pursuit trajectory as

\begin{widetext}
\begin{equation}
C_{\rm ij}(t,t') = \left[\begin{array}{c}{\bf v}_{\rm dir}^{({\rm i})} (t)  \\{\bf v}_{\rm speed}^{({\rm i})} (t)    \\{\bf v}_{\rm time}^{({\rm i})} (t)   \end{array}\right]^T
{\bf }\left[\begin{array}{ccc}\langle \delta \theta \delta \theta \rangle  &  \langle \delta \theta \delta v \rangle  &  \langle \delta \theta \delta t_0 \rangle  \\\langle \delta v \delta \theta \rangle   & \langle \delta v \delta v \rangle   &   \langle \delta v \delta t_0 \rangle\\\langle \delta t_0 \delta \theta \rangle  & \langle \delta t_0 \delta v \rangle  &   \langle \delta t_0 \delta t_0 \rangle\end{array}\right]{\bf }
\left[\begin{array}{c}{\bf v}_{\rm dir}^{({\rm j})} (t')  \\{\bf v}_{\rm speed}^{({\rm j})} (t')    \\{\bf v}_{\rm time}^{({\rm j})} (t')   \end{array}\right]
+ C^{\rm (back)}_{\rm ij}(t,t'),
\label{DeltaC-decomp}
\end{equation}  
\end{widetext}
\noindent where the terms $\langle\delta\theta\delta\theta\rangle$, $\langle\delta\theta\delta v\rangle$, etc.~are the covariances of the sensory errors, and we have abbreviated the partial derivative expressions for the modes of variation as  
${\bf v}_{\rm dir} \equiv {\partial \mathbf{v}/(t;t_0,v,\theta)}/{\partial \theta} $, ${\bf v}_{\rm speed} \equiv {\partial \mathbf{v}/(t;t_0,v,\theta)}/{\partial v}$, and ${\bf v}_{\rm time} \equiv {\partial \mathbf{v}/(t;t_0,v,\theta)}/{\partial t_0}$.  The fact that $C$ can be written in this form  implies not only that the variations in pursuit will be three dimensional, but that we can predict in advance what these dimensions should be.  Experimentally we find that
the three relevant dimensions have $96\%$ overlap with axes corresponding to ${\bf v}_{\rm dir}$, ${\bf v}_{\rm speed}$ and ${\bf v}_{\rm time}$.  

These results strongly support the hypothesis that the observable variations in motor output are dominated by the errors that the brain makes in estimating the parameters of its sensory inputs, as if the rest of the processing and motor control circuitry were effectively noiseless, or more precisely that they contribute only at the level of background variation in the movement.  Further, the magnitude and time course of noise in sensory estimation are comparable to the noise sources that limit perceptual discrimination \cite{osborne+al_05,osborne+al_07}.  This unexpected result challenges our intuition that noise in the execution of movement creates behavioral variation, and it forces us to consider that errors in sensory estimation may set the limit to behavioral precision.  Our findings are consistent with the idea that the brain can minimize the impact of noise in motor execution in a task specific manner \cite{harris+wolpert_98, todorov+jordan_02}, although it suggests a novel origin for that noise in the sensory system.  The precision of smooth pursuit fits well with the broader view that the nervous system can approach optimal performance at critical tasks \cite{todorov_04,bialek_02, bialek_87, barlow_81}.

 \subsection*{The way the worm wiggles}
 
 \begin{figure}[htb]
\includegraphics[width=0.95\columnwidth]{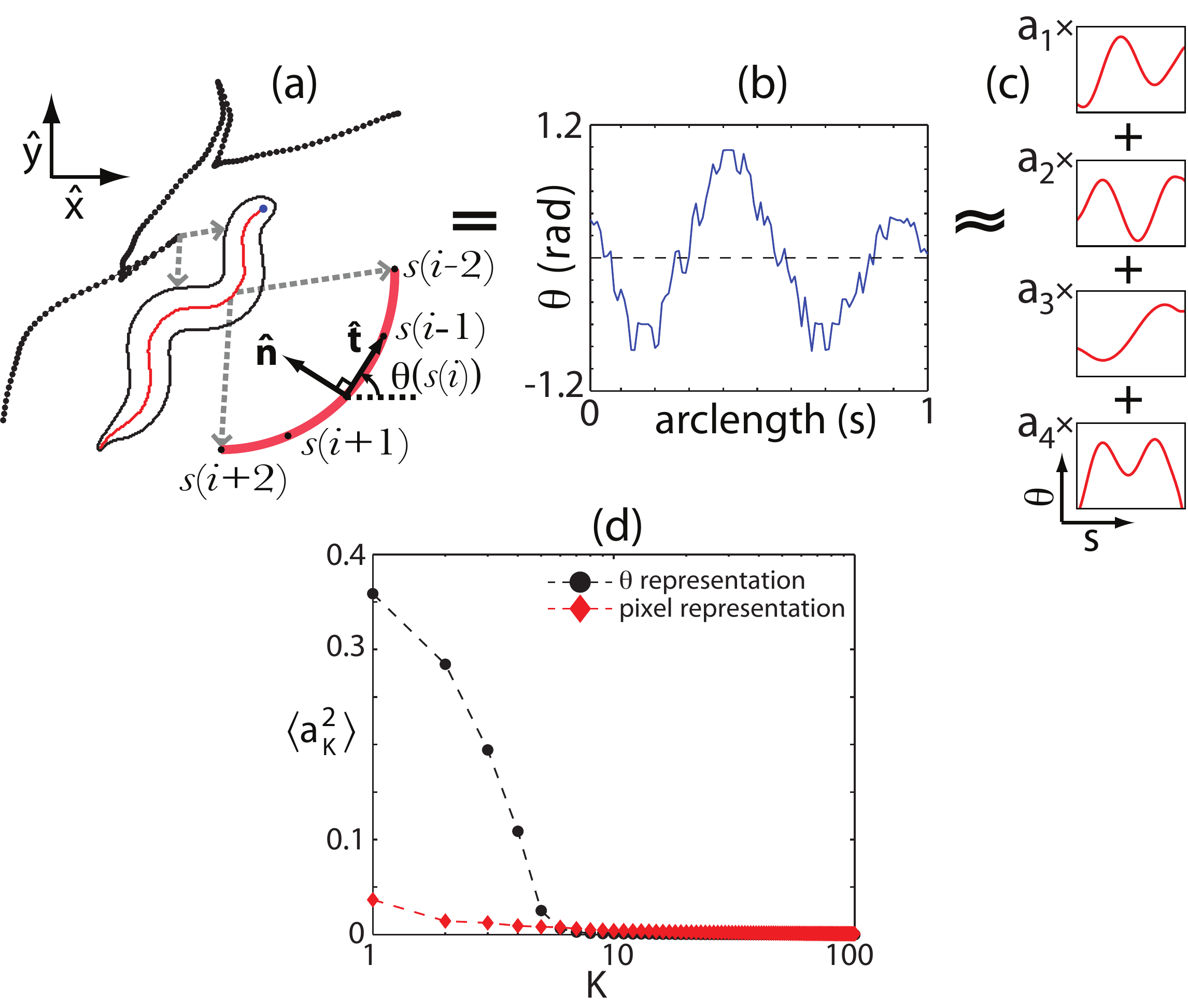}
\caption{The low-dimensional space of worm postures \cite{stephens+al_07}. (a) We use tracking video microscopy
to record images of the worm's body at high spatiotemporal resolution as it crawls along a flat agar surface.  Dotted lines trace the worm's centroid trajectory and the body outline and centerline skeleton are extracted from the microscope image on a single frame.
(b) We characterize worm shape by the tangent angle $\theta$ vs.~arc length $s$ of the centerline skeleton.
(c) We decompose each shape into four dominant modes by projecting $\theta(s)$ along the eigenvectors of the shape covariance matrix (eigenworms). (d, black circles) The fraction of total variance captured by each projection. The four eigenworms account for  $\sim 95\%$ of the variance within the space of shapes.
(d, red diamonds) The fraction of total variance captured when worm shapes are represented by images of the worm's body;
the low dimensionality is hidden in this pixel representation.}
\label{fig:fig2}
\end{figure}

The  free motion of the nematode {\em C. elegans}  on a flat agar plate provides an ideal opportunity to quantify 
the (reasonably) natural behavior of an entire organism \cite{stephens+al_07}. 
Under such conditions, changes in the worm's sinuous body shape support a variety of motor behaviors, including forward and backward crawling and large body bends known as $\Omega-$turns \cite{croll_75}.
Tracking microscopy provides high spatial and temporal resolution images of the worm over long periods of time, and from these images we can see that fluctuations  in the thickness of the worm are small, so most variations in the shape are captured by the curve that passes through the center of the body.  We measure position along this curve (arc length) by the variable $s$, normalized so that $s=0$ is the head and $s=1$ is the tail.  The position of the body element at $s$ is denoted by ${\bf x}(s)$,  but it is more natural to give an ``intrinsic'' description of this curve in terms of the  tangent angle  $\theta (s)$, removing our choice of coordinates by rotating each image so that the mean value of $\theta$ along the body always is zero.  Sampling at $N=100$ equally spaced points along the body,  each shape is described completely  by a $100-$dimensional vector (Fig 2a,b).

As we did with smooth pursuit eye movements, we seek a low dimensional space that underlies the shapes we observe.  In the simplest case, this space is a Euclidean projection of the original high dimensional space so that the covariance matrix of angles, $C(s, s' ) = \langle(\theta(s)-\langle\theta\rangle)( \theta (s')-\langle\theta\rangle)\rangle$, will have only a small number of significant eigenvalues.  For {\em C. elegans} this is exactly what we find, as shown in Fig 2c,d: over $95\%$ of the variance in body shape is accounted for by projections along just four dimensions (`eigenworms', red curves in Fig 2c).  Further, the trajectory in this low dimensional space of shapes predicts the motion of the worm over the agar surface \cite{stephens+al_09a}.
Importantly, the simplicity that we find depends on our choice of initial representation.  For example, if we take raw images of the worm's body, cropped to a minimum size ($300\times160$ pixels) and aligned to remove rigid translations and rotations, the variance across images is spread over hundreds of dimensions.    

The tangent angle representation and projections along the eigenworms provide a compact yet substantially complete description of worm behavior.  In distinction to previous work (see e.g.~\cite{croll_75,lockery+al_99,gray+al_05}), this description is naturally aligned to the organism,  fully computable from the video images with no human intervention, and also simple. In the next section we show how these coordinates can be also used to explore dynamical questions posed by the behavior of {\it C. elegans}.

\subsection*{Dynamics of worm behavior}
We have found low dimensional structure in the smooth pursuit eye movements of monkeys and in the free wiggling of nematodes.   Can this simplification inform our understanding of behavioral dynamics---the emergence of discrete behavioral states, and the transitions between them?  Here we  use the trajectories of {\em C. elegans} in the low dimensional space to construct an explicit stochastic model of crawling behavior, and then show how long-lived states and transitions between them emerge naturally from this model.

\begin{figure}[bt]
\includegraphics[width=0.95\columnwidth]{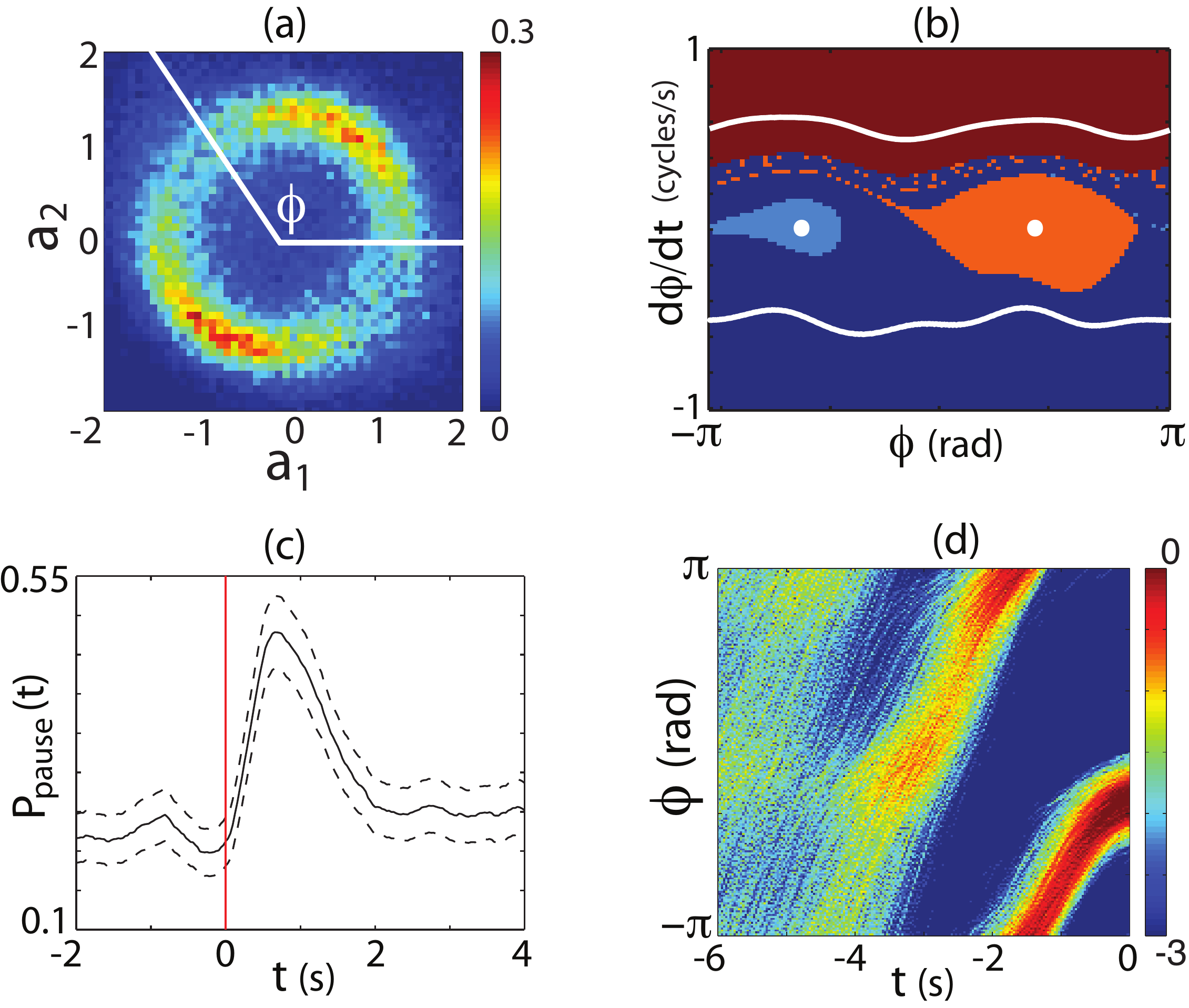}
\caption{Worm behavior in the eigenworm coordinates. 
(a) Amplitudes along the first two eigenworms oscillate, with nearly constant amplitude but time varying phase $\phi = \tan^{-1}(a_2 / a_1)$.  The shape coordinate $\phi(t)$ captures the phase of the locomotory wave moving along the worm's body. (b) The phase dynamics from Eq (\ref{langevin_omega}) reveals attracting trajectories in worm motion: forward and backward limit cycles (white lines), and two instantaneous pause states (white circles). Colors denote the basins of attraction for each attracting trajectory. (c) In an experiment in which the worm receives  a weak thermal impulse at time $t=0$, we use the basins of attraction of (b) to label the instantaneous state of the worm's behavior and compute the time dependent probability that a worm is in either of the two pause states.  The pause states uncover an early-time stereotyped response to the thermal impulse. (d) The probability density of the phase (plotted as $\log P(\phi |t)$),  illustrating stereotyped reversal trajectories consistent with a noise-induced transition from the forward state.  Trajectories were generated using Eq (\ref{langevin_omega}) and aligned to the moment of a spontaneous reversal at $t=0$. } 
\label{fig3}
\end{figure}

Of the four dimensions in shape space that characterize the crawling of {\em C. elegans}, motions along the first two combine to form an oscillation, corresponding to the  wave which passes along the worm's body and drives it forward or backward.  Here, we focus on the phase of this oscillation,  $\phi=\tan^{-1}{(a_2/a_1)}$ (Fig 3a), and construct, from the observed trajectories, a stochastic dynamical system,  analogous to the Langevin equation for a Brownian particle.
Since the worm can crawl both forward and backward, the phase dynamics is minimally a second order system,
\begin{eqnarray}
{d\phi  \over dt}&=& \omega  ,  \label{def_omega}  \nonumber \\
{d\omega \over dt}&=&F(\omega,\phi) +\sigma(\omega,\phi)\eta(t),
\label{langevin_omega}
\end{eqnarray}
where $\omega$ is the phase velocity and $\eta(t)$ is the noise---a random component of the phase acceleration not related to the current state of the worm---normalized so that $\langle \eta(t) \eta (t') \rangle = \delta (t-t')$.
As explained in Ref \cite{stephens+al_07}, we can recover the ``force'' $F(\omega,\phi)$ and the local noise strength $\sigma(\omega,\phi)$ from the raw data, so no further ``modeling'' is required.

Leaving aside the noise, Eq (\ref{langevin_omega})  describes a dynamical system in which there are multiple attracting trajectories (Fig 3b): two limit cycle attractors corresponding to forward and backward crawling (white lines) and two pause states (white circles) corresponding to an instantaneous freeze in the posture of the worm.  Thus, underneath the continuous, stochastic dynamics we find four discrete states which correspond to well defined classes of behavior.  We emphasize that these behavioral classes are emergent---there is nothing discrete about the phase time series $\phi(t)$, nor have we labelled the worm's motion by subjective criteria.   While forward and backward crawling are obvious behavioral states, the pauses are more subtle.  Exploring the worm's response to gentle thermal stimuli, one can see that there is a relatively high probability of a brief sojourn in one of the pause states (Fig \ref{fig3}c).  Thus, by identifying the attractors---and the natural time scales of transitions between them---we  uncover a more reliable component of the worm's response to sensory stimuli \cite{stephens+al_07}.

The noise term generates small fluctuations around the attracting trajectories, but more dramatically drives transitions among the attractors, and these transitions are predicted to occur with stereotyped trajectories \cite{stephens+al_09b}.    In particular, the Langevin dynamics in Eq (\ref{langevin_omega}) predict spontaneous transitions between the attractors that correspond to forward and backward motion.   To quantify this prediction, we run long simulations of the dynamics,  choose moments in time when the system is near the forward attractor ($0.1 < d\phi/dt < 0.6\,{\rm cycles/s}$), and then compute the probability that the trajectory has not reversed ($d\phi/dt < 0$) after a time $\tau$ following this moment.  If reversals are rare, this survival probability should decay exponentially, $P(\tau ) =  \exp(-\tau/\langle\tau\rangle )$, and this is what we see, with the predicted mean time to reverse $\langle\tau\rangle = 15.7\pm 2.1\,{\rm s}$, where the error reflects variations  across an ensemble of worms.

We next examine the real trajectories of the worms, performing the same analysis of reversals by measuring the survival probability in the forward crawling state.  We find that the data obey an exponential distribution, as predicted by the model, and the experimental mean time to reversal is $\langle\tau_{\rm data}\rangle = 16.3\pm 0.3\,{\rm s}$.   This observed reversal rate agrees with the model predictions within error bars, and this corresponds to a precision of $\sim 4\%$, which is quite surprising.  It should be remembered that we make our model of the dynamics by analyzing how the phase and phase velocity at the time $t$ evolve into phase and phase velocity at time $t+dt$, where the data determine $dt = 1/32\,{\rm  s}$.  Once we have the stochastic dynamics, we can use them to predict the behavior on long time time scales.   While we define our model on the timescale of a single video frame ($dt$), behavioral dynamics emerge that are nearly three orders of magnitude longer ($\langle\tau\rangle/dt \sim 500$), with no adjustable parameters \cite{stephens+al_09b}.  

In this model, reversals are noise driven transitions between attractors, in much the same way that  chemical reactions are  thermally driven transitions between attractors in the space of molecular structures \cite{hanggi+al_90}.  In the low noise limit, the trajectories that carry the system from one attractor to another  become stereotyped  \cite{dykman+al_94}.    Thus, the trajectories that allow the worm to escape from the forward crawling attractor are clustered around  prototypical trajectories, and this is seen both in the simulations (Fig \ref{fig3}d) and in the data \cite{stephens+al_09b}.  

In fact, many organisms, from bacteria to humans, exhibit discrete, stereotyped motor behaviors.  A common view is that these behaviors are stereotyped because they are triggered by specific commands, and in some cases we can even identify ``command neurons'' whose activity provides the trigger \cite{bullock}.  In the extreme, discreteness and stereotypy of the behavior reduces to the discreteness and stereotypy of the action potentials generated by the command neurons, as with the escape behaviors in fish triggered by spiking of the Mauthner cell \cite{mauthner}.  But the stereotypy of spikes itself emerges from the continuous dynamics of currents, voltages and ion channel populations \cite{hodgkin+huxley_52d,modHH}.  The success here of the stochastic phase model  in predicting the observed reversal characteristics of {\em C. elegans} demonstrates that stereotypy can also emerge directly from the dynamics of the behavior itself.

\section{Maximum entropy models of natural networks}
Much of what happens in living systems is the result of interactions among large networks of elements---many amino acids interact to determine the structure and function of proteins, many genes interact to define the fates and states of cells, many neurons interact to represent our perceptions and memories,  and so on.   Even if each element in a network achieves only two values, the number of possible states in a network of $N$ elements is $2^N$, which easily becomes larger than any realistic experiment (or lifetime!) can sample, the same dimensionality problem that we encountered in movement behavior. Indeed, a lookup table for the probability of finding a network in any one state has $\sim 2^N$ parameters, and this is a disaster.  To make progress we search for a simpler class of models with {\em many} fewer parameters.  

We seek an analysis of living networks that leverages increasingly high-throughput experimental methods such as the recording from large numbers of neurons simultaneously.  These experiments provide, for example, reliable information about the correlations between the action potentials generated by pairs of neurons.  In a similar spirit, we can measure the correlations between amino acid substitutions at different sites across large families of proteins. Can we use these pairwise correlations to say anything about the network as a whole?  While there are an infinite number of models that can generate a given pattern of pairwise correlations, there is a {\em unique} model that reproduces the measured correlations and adds no additional structure.  This minimally structured model is the one that maximizes the entropy of the system  \cite{jaynes_57}, in the same way that the thermal equilibrium (Boltzmann) distribution maximizes the entropy of a physical system given that we know its average energy.  

\subsection*{Letters in words}

To see how the maximum entropy idea works, we examine an example where we have some intuition for the states of the network.  Consider the spelling of four letter English words \cite{stephens+bialek_08}, where at positions ${\rm i} = 1, 2, 3, 4$ in the word we can chose a variable $x_{\rm i}$ from 26 possible values.  A word is then represented by the combination ${\bf x} \equiv \{x_1, x_2, x_3, x_4\}$, and we can sample the distribution of words, $P({\bf x})$, by looking through a large corpus of writings, for example the collected novels of Jane Austen \cite{letters_notes}.  If we don't know anything about the distribution of states in this network, we can maximize the entropy of the distribution  $P({\bf x})$ by having all possible combinations of letters be equally likely, and then the entropy is $S_0 =-\sum P_0 \log_2{P_0}=4\times \log_2 (26) = 18.8\,{\rm bits}$.  But, in actual English words, not all letters occur equally often, and this bias in the use of letters is different at different positions in the word.  If we take these ``one letter'' statistics into account, the maximum entropy distribution is the independent model, 
\begin{equation}
P^{(1)}({\bf x}) = P_1 (x_1) P_2(x_2) P_3(x_3) P_4 (x_4),
\label{P1}
\end{equation}
where $P_{\rm i}(x)$ is the easily-measured probability of finding letter $x$ in position $\rm i$.  Taking account of actual letter frequencies lowers the entropy to $S_1 = 14.083\pm 0.001\,{\rm bits}$ where the small error bar is derived from sampling across the $\sim 10^6$ word corpus.

The independent letter model defined by $P^{(1)}$ is clearly wrong:  the most likely words are `thae', `thee' and `teae.'    Can we build a better approximation to the distribution of words by including correlations between pairs of letters?  The difficulty is that now there is no simple formula like Eq (\ref{P1}) which connects the maximum entropy distribution for $\bf x$ to the measured distributions of letter pairs $(x_{\rm i}, x_{\rm j})$.  Instead we know analytically the form of the distribution,
\begin{equation}
P^{(2)}({\bf x})  = {1\over Z} \exp\left[
-\sum_{{\rm i} > {\rm j}} V_{\rm ij} (x_{\rm i} , x_{\rm j}) 
\right],
\label{boltz}
\end{equation}
where all of the coefficients $V_{\rm ij} (x , x')$ have to be chosen to reproduce the observed correlations between pairs of letters.  This is complicated, but much less complicated than it could be---by matching all the pairwise correlations we are fixing $\sim 6\times (26)^2$ parameters, which is vastly smaller than the $(26)^4$ possible combinations of letters.

The model in Eq (\ref{boltz}) has exactly the form of the Boltzmann distribution for a physical system in thermal equilibrium, where the letters ``interact'' through a potential energy $V_{\rm ij} (x , x')$.  The essential simplification is that there are no explicit interactions among triplets or quadruplets---all the higher order correlations must be consequences of the pairwise interactions.   We know that in many physical systems this is a good approximation,  that is $P \approx P^{(2)}$.  However, the rules of spelling (e.g., i before e except after c) seem to be in explicit conflict with such a simplification.  Nonetheless, when we apply the model in Eq (\ref{boltz}) to English words, we find reasonable phonetic constructions.  
Here we leave aside the problem of how one finds the potentials $V_{\rm ij}$ from the measured correlations among pairs of letters (see Refs \cite{darroch+ratcliff_72,tkacik+al_06,broderick+al_07,gammor+al_09,sessak+monasson_09,mezard+mora_09,roudi+al_09b}), and discuss the results.

Once we construct a maximum entropy model of words using Eq (\ref{boltz}), we find  that the entropy of the pairwise model is $S_2  = 7.471\pm 0.006\,{\rm bits}$, about half the entropy of independent letters $S_1$.  A rough way to think about this result is that if letters were chosen independently, there would be $2^{S_1}\sim 17,350$ possible four letter words.  Taking account of the pairwise correlations reduces this vocabulary by a factor of $2^{S_1 - S_2}\sim 100$, down to effectively $\sim 178$ words. In fact, the Jane Austen corpus is large enough that we can estimate the true entropy of the distribution of four letter words, and this is $S_{\rm full} = 6.92\pm 0.003\,{\rm bits}$. Thus the pairwise model captures $\sim 92\%$ of the entropy reduction relative to choosing letters independently, and hence accounts for almost all of the restriction in vocabulary provided by the spelling rules and the varying frequencies of word usage.  The same result is obtained with other corpora, so this is not a peculiarity of an author's style.  

\begin{figure}[t]
\includegraphics[width=0.95\columnwidth]{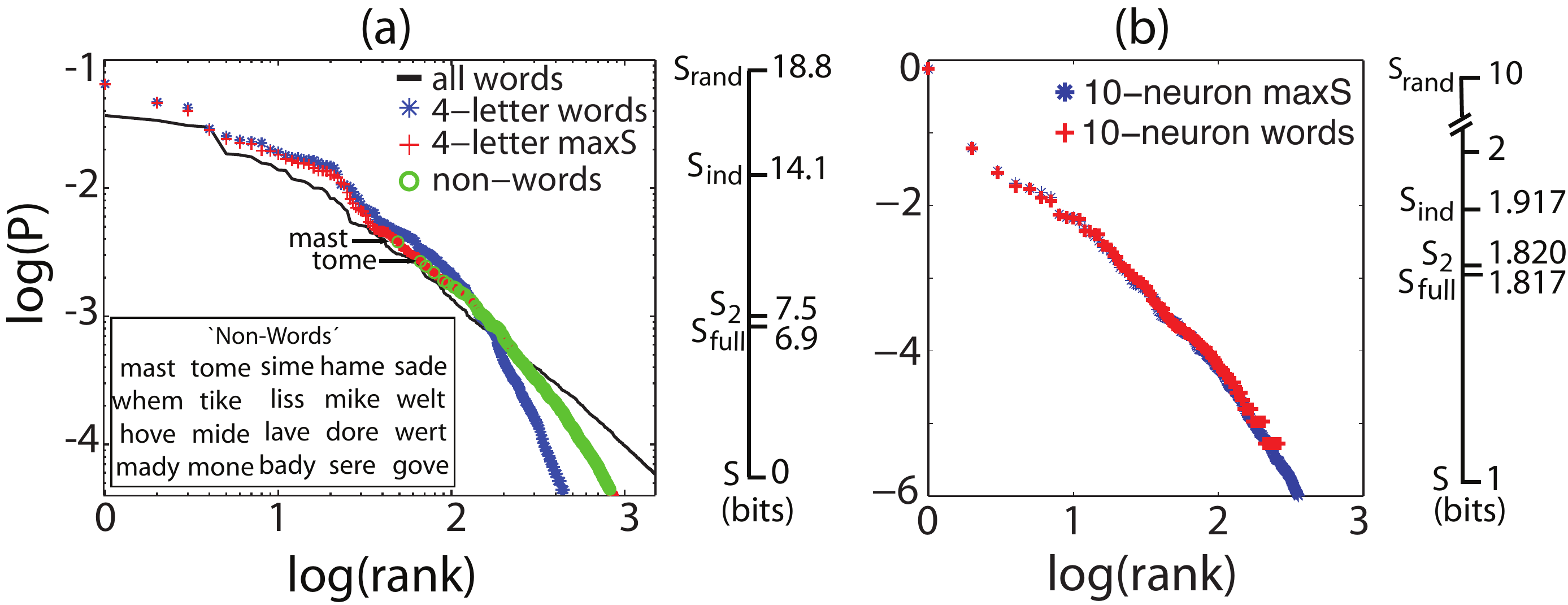}
\caption{For networks of neurons and letters, the pairwise maximum entropy model provides an excellent approximation to the probability of network states.  In each case, we show the  Zipf plot for real data (blue) compared to the pairwise maximum entropy approximation (red).  Scale bars to the right of each plot indicate the entropy captured by  the pairwise model. (a) Letters within four letter English words \cite{stephens+bialek_08}.  The maximum entropy model also produces `non-words'  (inset, green circles) that never appeared in the full corpus but nonetheless contain realistic phonetic structure.   (b) 10 neuron patterns  of spiking and silence in the vertebrate retina \cite{schneidman+al_06}. }
\label{fig:fig4}
\end{figure}

We can look more closely at the predictions of the maximum entropy model in a ``Zipf plot,'' ranking the words by their probability of occurrence  and plotting probability vs. rank, as in Fig \ref{fig:fig4}.  The predicted Zipf plot almost perfectly overlays what we obtain by sampling the corpus, although some weight is predicted to occur in words that do not appear in Austen's writing.  Many of these are real words that she happened not to use, and others are perfectly pronounceable English even if they are not actually words.  Thus, despite  the complexity of spelling rules, the pairwise model  captures a very large fraction of the structure in the network of letters.

\subsection*{Spiking and silence in neural networks}
Maximum entropy models also provide a good approximation to the patterns of  spiking 
in the neural network of the retina.  In a network of neurons where the variable $x_{\rm i}$ marks the presence ($x_{\rm i} = +1$) or absence  ($x_{\rm i} = -1$) of an action potential from neuron $\rm i$ in a small window of time, the state of the whole network is given by the pattern of spiking and silence across the entire population of neurons, ${\bf x} \equiv \{x_1, x_2, \cdots , x_N\}$.    
In the original example of these ideas, Schneidman et al  \cite{schneidman+al_06} looked at groups of $N=10$ nearby neurons in the vertebrate retina as it responded to naturalistic stimuli, with the results shown in Fig \ref{fig:fig4}.  Again we see that the pairwise model does an excellent job, capturing $\sim 90\%$ or more of the reduction in entropy, reproducing the  Zipf plot, and even predicting the wildly varying probabilities of the particular patterns of spiking and silence (see Fig 2a of Ref \cite{schneidman+al_06}). 

The maximum entropy models discussed here are important because they often capture a large fraction of the interactions present in natural networks while simultaneously avoiding a combinatorial explosion in the number of parameters. This is true even in cases where interactions are strong enough so that independent  (i.e.~zero neuron-neuron correlation) models fail dramatically. Such an approach has also recently been used to show how network functions such as stimulus decorrelation and error correction reflect a trade-off between efficient consumption of finite neural bandwidth and the use of redundancy to mitigate noise \cite{tkacik+al_10}. 

As we look at larger networks, we can no longer compute the full distribution and thus we cannot directly compare the full entropy with it's pairwise approximation.  We can, however, check many other predictions and the maximum entropy model works well, at least to $N=40$ \cite{tkacik+al_06,tkacik+al_09}.   
Related ideas have also been applied to a  variety of neural networks with similar findings \cite{shlens+al_06,tang+al_08,yu+al_08,shlens+al_09} (however, also see \cite{other_neurons_2} for differences), which suggest that the networks in the retina are typical of a larger class of natural ensembles.

\subsection*{Metastable states} 

As we have emphasized in discussing Eq (\ref{boltz}),  maximum entropy models are {\em exactly} equivalent to Boltzmann distributions, and thus define an effective ``energy'' for each possible configuration of the network.  States of high probability correspond to low energy, and we can think of an ``energy landscape'' over the space of possible states, in the spirit of the Hopfield model for neural networks \cite{hopfield_82}.    Once we construct this landscape, it is clear that some states are special because they sit at the bottom of a valley---at local minima of the energy.  For networks of neurons, these special states are such that flipping any single bit in the pattern of spiking and silence across the population generates a state with lower probability.  For words, a local minimum of the energy means that changing any one letter produces a word of lower probability.    

\begin{figure}[htb]
\includegraphics[width=0.95\columnwidth]{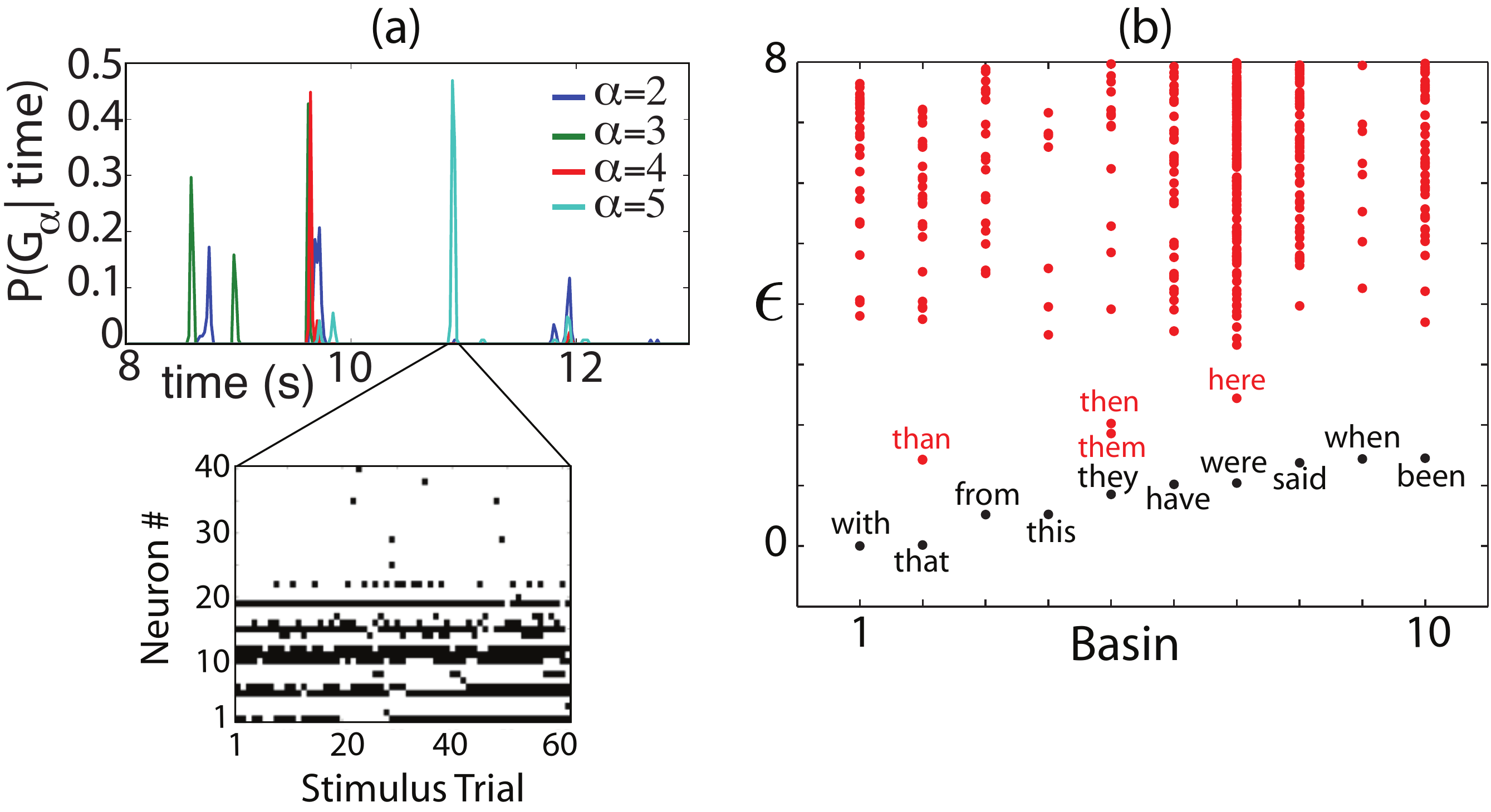}
\caption{Metastable states in the energy landscape of networks of neurons and letters.   (a) Probability that the 40 neuron system is found within the basin of attraction of each nontrivial locally stable state $G_{\alpha}$ as a function of time during the 145 repetitions of the stimulus movie.  The inset shows the state of the entire network at the moment it enters the basin of $G_5$, on 60 successive trials.  (b) The energy landscape ($\epsilon = -\ln P$) in the maximum entropy model of letters in words.  We order the basins in the landscape by decreasing probability of their ground states, and show  the low energy excitations in each basin.}
\label{fig:fig5}
\end{figure}

 The picture of an energy landscape on the states of a network may seem abstract, but the local minima can (sometimes surprisingly) have functional meaning, as shown in Fig \ref{fig:fig5}.  In the case of the retina, a maximum entropy model was constructed to describe  the states of spiking and silence in a population of $N=40$ neurons as they respond to naturalistic inputs, and this model predicts the existence of several non--trivial local minima  \cite{tkacik+al_06,tkacik+al_09}.  Importantly, this analysis does not make any reference to the visual stimulus.  But if we play the same stimulus movie many times, we see that the system returns to the same valleys or basins surrounding these special states, even though the precise pattern of spiking and silence is not reproduced from trial to trial (Fig \ref{fig:fig5}a).  This suggests that the response of the population can be summarized by which valley the system is in, with the detailed spiking pattern being akin to variations in spelling.  To reinforce this analogy, we can look at the local minima of the energy landscape for four letter words.
 
In the maximum entropy model for letters, we find 136 of  local minima, of which the 10 most likely are shown in Fig 5b.  More than 2/3 of the entropy in the full distribution of words is contained in the distribution over these valleys, and in most of these valleys there is a large gap between the bottom of the basin (the most likely word) and the next most likely word. Thus, the entropy of the letter distribution is dominated by states which are not connected to each other by single letter substitutions, perhaps reflecting a pressure within language to communicate without confusion.

\section{Discussion}

Understanding a complex system necessarily involves some sort of simplification.   We have emphasized that, with the right data, there are mathematical methods which allow a system to ``tell us'' what sort of simplification is likely to be useful.

Dimensionality reduction is perhaps the most obvious method of simplification---a direct reduction in the number of variables that we need to describe the system.  The examples of {\em C. elegans} crawling and smooth pursuit eye movements are compelling because the reduction is so complete, with just three or four coordinates capturing $\sim 95\%$ of all the variance in behavior.  In each case, the low dimensionality of our description provides functional insight, whether into origins of stereotypy or the possibility of optimal performance.    The idea of dimensionality reduction in fact has a long history in neuroscience, since receptive fields and feature selectivity are naturally formalized by saying that neurons are sensitive only to a limited number of dimensions in stimulus space \cite{spikes,bialek+ruyter_05,mid,schwartz+al_06}.  More recently it has been emphasized that quantitative models of protein/DNA interactions are equivalent to the hypothesis that proteins are sensitive only to limited number of dimensions in sequence space \cite{kinney+al_07,kinney+al_10}.  

The maximum entropy approach achieves  a similar simplification for networks; it searches for simplification not in the number of variables, but in the number of possible interactions among these variables.  The example of letters in words shows how this simplification retains the power to describe seemingly combinatorial patterns.  For both neurons and letters, the mapping of the maximum entropy model onto an energy landscape  points to special states of the system that seem to have functional significance.    There is an independent stream of work which emphasizes the sufficiency of pairwise correlations among amino acid substitutions in defining functional families of proteins \cite{socolich+al_05,russ+al_05}, and this is equivalent to the maximum entropy approach \cite{bialek+ranganathan_07}; explicit construction of the maximum entropy models for antibody diversity again points to the functional importance of the metastable states \cite{mora+al_10}.  

Although we have phrased the ideas of this paper essentially as methods of data analysis, the repeated successes of mathematically equivalent models (dimensionality reduction in movement and maximum
entropy in networks) encourages us to seek unifying theoretical principles that give rise to behavioral 
simplicity.   Finding such a theory, however, will only be possible if we observe behavior in sufficiently unconstrained contexts so that simplicity is something we discover rather than impose.

\bigskip

\begin{acknowledgments}
We are grateful to DW Pfaff and his colleagues for organizing the Sackler Colloquium, and for providing us the opportunity to bring together several strands of thought.  We thank our many collaborators who have worked with us on these ideas, and made it so much fun:  MJ Berry II, CG Callan, B Johnson--Kerner, SG Lisberger, T Mora, SE Palmer, R Ranganathan, WS Ryu, E Schneidman, R Segev, S Still, G Tka\v{c}ik and A Walczak.  This work has been supported in part by grants from the National Science Foundation, the National Institutes of Health, and the Swartz Foundation.
\end{acknowledgments}

\end{document}